\documentclass{aa}
\usepackage{txfonts}

\usepackage{natbib}
\usepackage{graphicx}
\usepackage[english]{babel}
\usepackage{lscape}

\newcommand{\gppr}{\stackrel{>}{\scriptstyle \sim}}
\newcommand{\gappr}{\raisebox{-0.4ex}{$\gppr$}}
\newcommand{\lppr}{\stackrel{<}{\scriptstyle \sim}}
\newcommand{\lappr}{\raisebox{-0.4ex}{$\lppr$}}

\newcommand{\Porb}{\mbox{$P_\mathrm{orb}$}}

\newcommand{\Mwd}{\mbox{$M_\mathrm{wd}$}}
\newcommand{\Msec}{\mbox{$M_\mathrm{sec}$}}

\begin{document}

\title{The statistical significance of the superhump signal in U\,Gem}
\authorrunning{M.R. Schreiber}
\titlerunning{Superhumps in U\,Gem}
\author{Matthias R. Schreiber\inst{1,2}}
\institute{
Departamento de F\'isica y Astronom\'ia, Facultad de Ciencias, Universidad 
de Valpara\'iso, Valpara\'iso, Chile 
\and
Astrophysikalisches Institut Potsdam, An der Sternwarte 16, D-14482
Potsdam\\
\email{Matthias.Schreiber@uv.cl}
}
\offprints{M.R. Schreiber}

\date{Received / Accepted }

\abstract
{Although its well determined mass ratio of $q=\Msec/\Mwd=0.357\pm0.007$ 
should avoid superoutbursts according to the thermal tidal instability 
model, the prototypical dwarf nova U\,Gem experienced in 1985 an 
extraordinary long outburst resembling very much superoutbursts observed in 
SU\,UMa systems. Recently, the situation for the model became even worse 
as superhump detections have been reported for the 1985 outburst of U\,Gem.}
{The superhump signal is noisy and the evidence provided by simple
periodograms seems to be weak. Therefore and because of the importance
for our understanding of superoutbursts and superhumps, we determine 
the statistical significance of the recently published detection of 
superhumps in the AAVSO light curve of the famous long 1985 outburst 
of U\,Gem.}
{Using Lomb-Scargle periodograms, analysis of variance (AoV), 
and Monte-Carlo methods we analyse the 
160 visual magnitudes obtained by the AAVSO during the outburst and relate our 
analyse to previous superhump detections.}
{The 160 data points of the outburst alone do not contain a statistically
significant period. However, using additionally the characteristics of 
superhumps detected previously in other SU\,UMa systems and searching only for
signals that are consistent with these, we derive a $2\sigma$ 
significance for the superhump signal. The alleged appearance of an 
additional superhump at the end of the outbursts appears to be statistically 
insignificant. }
{Although of weak statistical significance, the superhump signal of the 
long 1985 outburst of U\,Gem can be interpreted as further indication for the
SU\,UMa nature of this outburst. This further contradicts the tidal 
instability model as the explanation for the superhump phenomenon.}


\maketitle

\section{Introduction}
Dwarf novae are non-magnetic CVs showing
quasi-regular outbursts, i.e. increased visual brightness of 2-5\,mag 
for several days which reappear typically on timescales of weeks to months
\citep[e.g.][for a review]{warner95-1}.
SU\,UMa stars are short-period, i.e. $\Porb\leq\,2.2$\,hr
dwarf novae whose light
curves consist of two types of outburst: normal dwarf nova
outbursts and superoutbursts which are
5-10 times longer as well as $\sim0.7$\,mag brighter. 
These superoutbursts also show pronounced humps
(called superhumps) reappearing with periods usually a few percent longer 
than the orbital one. 
The phenomenon is usually explained by tidal disc deformations
when the radius of the disc reaches the 3:1 resonance radius 
\citep{whitehurst88-1,whitehurst+king91-1,lubow91-1}.
This resonance is possible only if the mass ratio of the components
is small, i.e. $q\equiv\Msec/\Mwd\leq\,0.33$. Therefore the observed
appearance of superhumps in systems with short orbital periods and small mass
ratios is in general agreement with the tidal instability 
explanation for {\em superhumps}.
In contrast, there have been two possible scenarios for the triggering
mechanism of {\em superoutbursts} proposed: 
either they are also caused by the 3:1 resonance as 
it is assumed in the thermal tidal instability (TTI) model 
\citep[see e.g.][for a review]{osaki96-1}
or they are triggered by enhanced mass transfer (EMT) as proposed
by \citet[][]{vogt83-2} and \citet{smak84-1}.
According to the disc instability model the EMT scenario appears to be 
more plausible \citep{schreiberetal04-1}. On the other hand, 
the mechanism claimed to 
cause the mass transfer enhancement, i.e. irradiation of the secondary by 
the white dwarf and the boundary layer is not well understood 
\citep[see][for recent arguments on the EMT and the TTI models]{osaki+meyer03-1,smak04-1,osaki+meyer04-1,smak04-2,schreiberetal04-1,truss05-1}. 

\begin{figure}
\includegraphics[width=4.2cm, angle=270]{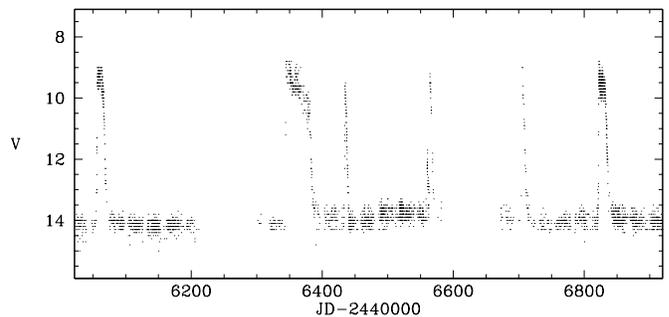}
\caption{\label{f-obs} Snapshot of the visual light curve of
U\,Gem including the extremely long 
``superoutburst'' (starting at $JD\sim\,2446344$) for which
\citet{smak+waagen04-1} reported the detection of superhumps. 
The data has been provided by the AAVSO.}
\end{figure}

The prototypical dwarf nova system U\,Gem plays  
a key role in the context of the discussion about superoutbursts,
superhumps, the EMT, and the TTI model: 
Its orbital period
is $4.25$\,hrs and its mass ratio of $q=0.357\pm0.007$ \citep{nayloretal05-1} 
is above the limit for tidal instabilities. U\,Gem normally shows 
regular dwarf nova outbursts with $10-20$\,days duration and 
recurrence times of $\sim100$\,days typical for long orbital period dwarf
novae. In 1985 U\,Gem exhibited a famous
long (45\,days) and large amplitude ($\sim\,0.5$\,mag brighter) outburst 
reminiscent of a SU\,UMa superoutburst 
(see Fig.\,1 or \citet[][]{masonetal88-1}).
According to the tidal instability model, 
the mass ratio of U\,Gem should prevent the appearance of superoutbursts.
Therefore, the 1985 outburst may indicate that 
superoutbursts are not caused by tidal instabilities
\citep[e.g.][]{lasota01-1}. 

Recently the situation became even more difficult for the TTI model: 
\citet{smak+waagen04-1} reported the detection of superhumps 
in the AAVSO light curve of the 1985 outburst of U\,Gem which 
- if true - has extremely far reaching consequences. 
Superhumps in U\,Gem would 
{\bf{(1.)}} contradict the general explanation for the superhump 
phenomenon and
{\bf{(2.)}} synchronise the simultaneous appearance of superhumps and
superoutbursts. The detection of superhumps in U\,Gem
would hence require to develop a new theory for superhumps and superoutbursts
which should not rely on tidal forces \citep[see
also][]{hameury+lasota05-1,hameury+lasota05-2}. Such a new scenario 
would be in agreement with the findings of \citet{kornet+rozyczka00-1} 
whose hydrodynamic TTI models do not predict superhumps 
if the full energy equation is taken into account. 
However, before completely abandon the thermal-tidal instability model 
one should be aware that determining periods in uneven 
datasets is a non-trivial statistical exercise. 
In particular, it is difficult to estimate the
significance of a signal in a periodogram. 
\citet{smak+waagen04-1} list many arguments why they {\em believe} in the
reality of the superhump periodicity ranging from the coherence of the 
period changes (their Fig.\,4) to the fact that the
amplitude of the alleged superhump signal is typical for this phenomenon.   
However, the signal shown in the 
periodograms is extremely weak and has been taken with some scepticism 
\citep{pattersonetal05-1}.
The claimed detection of superhumps in U\,Gem 
is of outstanding importance for both our understanding of the superhump 
phenomenon itself as well as the triggering mechanism of superoutbursts. 
Therefore we want to qualify the {\em{believe}} and the {\em{scepticism}} 
and present here a detailed analysis of the AAVSO data using not only 
discrete Fourier transforms but also analysis of variances (AoV)
and randomisation 
techniques to determine the statistical significance of the alleged 
periodicities. 
\begin{figure}
\includegraphics[width=8.cm, angle=0]{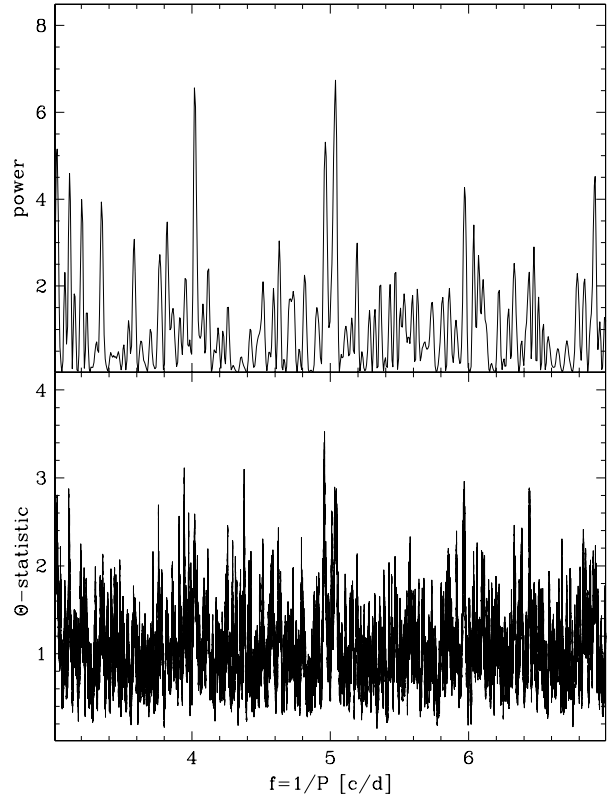}
\caption{\label{fig_ftaovpd} 
Periodograms for the AAVSO data of the long and bright 1985 outburst of
U\,Gem using the Lomb-Scargle algorithm and the AoV method. 
The power-spectrum and the analysis of variance (AoV)
statistic are calculated for $1\times10^5$ frequencies. The binning in the
case of the AoV periodogram is $N_c=2, N_b=10$. The double peaked 
shape of the claimed superhump signal around $f\sim5$\,c/d indicates that 
the period may be variable in time.
}
\end{figure}

\section{Periodograms}

Having selected the 160 AAVSO measurements representing 
the plateau of the long 1985 outburst of U\,Gem ($JD=2446344.66-2446381.43$),
we first analyse the data following \citet{smak+waagen04-1}: 
we subtract the linear
trend and calculate a simple Lomb-Scargle periodogram according to  
a classical discrete Fourier transform to construct a power spectrum. 
In addition, it is common and useful to analyse uneven time series 
of data using the method of phase dispersion minimisation (PDM)
\citep{stellingwerf78-1} or AoV \citep{schwarzenberg-czerny89-1}. 
Fig.\,\ref{fig_ftaovpd} shows periodograms obtained with the Lomb-Scargle
algorithm and the AoV method. Indeed, there is a peak at a 
frequency $f\sim5$\,c/d in the periodograms as reported 
by \citet{smak+waagen04-1} 
and the double peak shape indicates that the period could vary with
time.
We here repeat the question asked by these authors:
``Does this signal represent a real periodicity?''
Unfortunately, answering this question, i.e. determining the 
statistical significance of 
period detections in uneven datasets is not straight forward. 
In fact, a perfect analytical solution of the problem does not yet exist.
To determine the significance for time dependent signals and/or one 
particular time spacing of the data points, one can either use numerical 
simulations or semi-analytical approximations.

\section{Significance tests}

\begin{figure}
\includegraphics[width=8.5cm, angle=0]{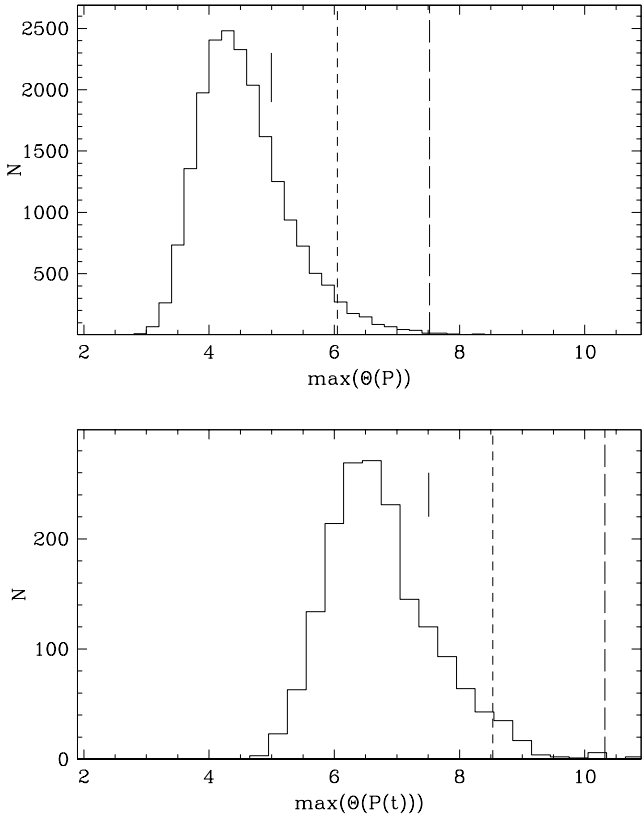}
\caption{\label{fig_histb} 
Distribution of the maximum of the $\Theta$-statistic for 5 bins and
using a broad range of trial periods, i.e. $3$\,c/d$\leq\,f\,\leq\,7$\,c/d. 
In the top panel we used only
constant periods while we took into account coherent time evolution for the
superhump periods in the bottom panel. 
The short solid vertical lines indicate the positions of the
claimed superhump signal. The dashed lines represent the $95\%$ and $99.7\%$ 
($3\,\sigma$) significance levels.  
In both cases the statistical significance of the superhump signal is
far from reaching $2\sigma$, i.e $p=0.275\pm0.003$ (top panel) and
$p=0.224\pm0.010$ (bottom panel).  
}
\end{figure}

Using the Lomb-Scargle formalism one can determine the significance using 
estimates for the number of independent frequencies $M$. 
The false alarm probability is then given by 
\begin{equation}
p(>z)\equiv1-(1-e^{-z})^M
\end{equation}
where $z$ is the normalised power of the most significant peak in the
periodogram \citep[see][]{pressetal92-1}.
In general $M$ depends on the frequency bandwidth, the
number of data points, and their detailed time spacing.
While \citet{horne+baliunas86-1} performed extensive Monte-Carlo 
simulations to determine $M$, \citet[][]{pressetal92-1} argue that one does 
not need to know $M$ very precisely
and that it usually should not be essentially different from the number of 
data points $N_0$. 
More recently, \citet[][his Sect. 3.2]{paltani04-1} presented 
an interesting and new semi-analytical method to estimate $M$. As this
formalism requires less extensive numerical simulations than the
\citet{horne+baliunas86-1} approach, it certainly
represents a promising new method. 

\begin{figure*}
\includegraphics[width=8.45cm, angle=0]{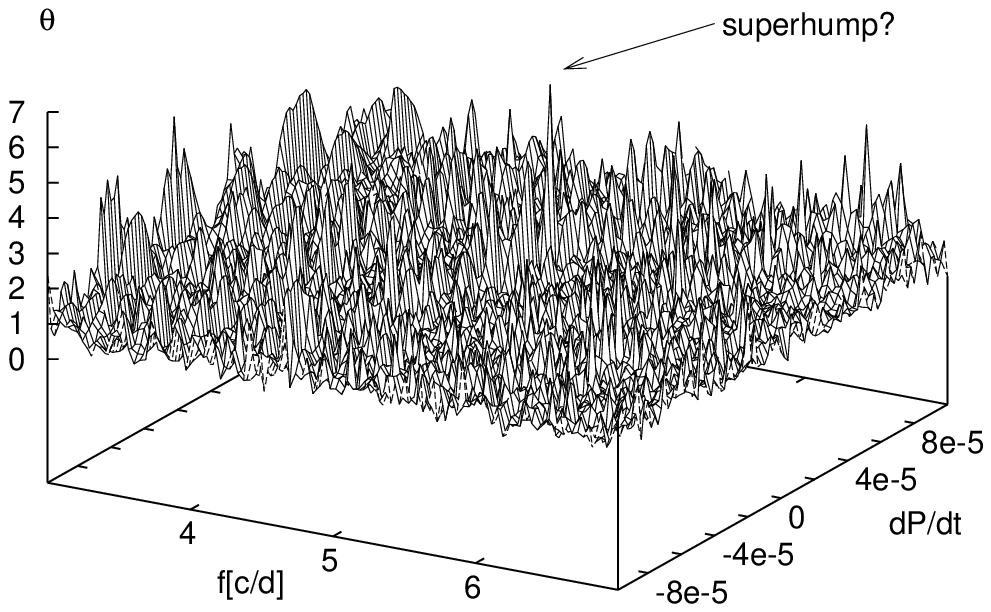}
\includegraphics[width=8.45cm, angle=0]{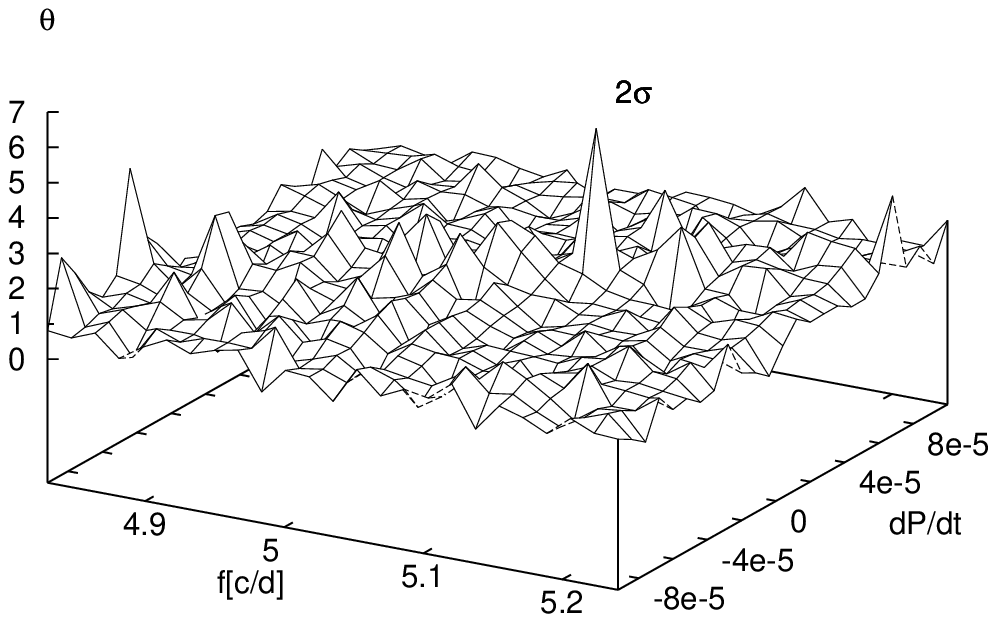}
\caption{\label{fig_fake}
Low resolution AoV periodograms taking into account a range of constant 
period derivatives illustrating the effect of restricting 
the range of trail periods. 
The periodograms are calculated using $N_b=5$ and $N_c=2$. We 
restricted our analysis to constant period
derivatives, i.e. we used $P(t)=P_0+a\,\dot{P}$ 
were $-0.0001\leq\,a\leq\,0.0001$. 
The left panel shows the periodogram for a broad rang of trial periods 
i.e. $1/3$\,d$\geq\,P_0\,\geq1/7$\,d.  
The peak at $f\sim5$\,c/d and $dP/dt\sim\,2e-5$ is hardly distinguishable 
from the noise and is not statistically significant. 
However, if we search only for periods consistent with the $\epsilon-\Porb$
relation, i.e. $4.83$\,c/d$\leq\,f\leq\,5.24$\,c/d 
(see Fig.\,\ref{fig_eps}), the peak produced by the superhump signal is
clearly the highest peak and we derive a $2\sigma$ significance. 
Please note, the data used to construct the figures above has been calculated
with a much lower resolution in $f$ and $dP/dt$ than used
in our detailed Monte-Carlo simulations.
%For this reason, the peaks are
%slightly smaller than   
%
% right panel four periodograms seem to represent the same noise with some 
%weak peaks caused by chance. However, as the periodograms are calculated using
%only constant trial periods and the range of plotted periods is rather large,
%this does not confirm the statistical insignificance of the superhump signal
%claimed by \citet{smak+waagen04-1}.   
}
\end{figure*}

An alternative to the Lomb-Scargle method are PDM and AoV.  
In both cases we know the probability distribution of the 
statistic $\Theta(P)$ if the 
period (P) is known (i.e. the beta and the Fisher-Snedecor (F) 
distribution
\citep{schwarzenberg-czerny97-1,schwarzenberg-czerny89-1}).
Unfortunately, 
replacing in Eq.\,(1) $e^{-z}$ with the corresponding beta or F 
probabilities leads to similar problems as above, i.e. one needs to estimate
the number of independent frequencies $M$
\citep[see also][]{hecketal85-1}. 
Instead \citet{nemec+nemec85-1} proposed a Fisher randomisation 
technique to numerically determine
the required probability distribution of the maximum of the statistic in the
considered period range $V$, i.e. $max(\Theta(P))$ for $P\in V$. 
Assuming that there is no periodicity in the data, the observations 
should be independent of the observing times and randomly redistributing 
the measurements should not give significantly different results. 
Generally speaking, 
the null hypothesis, i.e. assuming that the variations in the data 
represent just noise, is true 
if the original periodogram does not contain a particularly strong peak when
compared to those obtained after having randomly redistributed the 
observations. 
More specifically the method is as follows. First one calculates periodograms 
and the maximum of the statistic ($max(\Theta(P))$) 
for the observations using the PDM or AoV technique.
Then, this procedure is repeated many times after having each time 
randomly redistributed the measurements over the fixed times of 
observations. 
This Monte-Carlo experiment results in a distribution of maxima 
of the $\Theta$-statistic. 
The proportion of randomised datasets producing a value of 
$max(\Theta(P))$ which is equal to or 
larger than the value obtained for the observations gives the 
probability value $p$ (often also called the probability of error).  
The statistical significance of the detected signal is $1-p$.  
The error of $p$ obviously depends on the number of 
randomised datasets $n$, on the resolution in frequency, and the binning.   
Following \citet{nemec+nemec85-1} the standard error 
of $p$ can be approximated as $\sigma_p\sim[p(1-p)/n]^{1/2}$ and 
the $95$\% confidence level can be written as $p\pm\,2\sigma_p$. 

In the following Sections we use AoV and the just described 
randomisation method to determine the
statistical significance of the superhump signal in the light curve of the
1985 outburst of U\,Gem. 
To assure that our results are independent on the 
resolution we increased the number of frequencies until the obtained value 
of $p$ remained constant. We find that calculating $\Theta(P)$ for 
$10^5$ equally spaced frequencies in the range of  
$3\leq f \leq\,7$ is sufficient to exclude artefacts resulting from low 
resolutions. In addition, we performed several 
randomisation tests to estimate the influence 
of different binnings. The binning parameters we used 
are $N_b$ and $N_c$ as defined in \citet{stellingwerf78-1}. 
It turns out that $p$ is somewhat depending on the binning as long as
$N_c=1$ but remains more or less constant for $N_c\geq2$.  
Finally, we want to stress that although often very useful, 
Monte-Carlo or randomisation techniques have their limitations. 
Most importantly, they rely on the assumption of white noise, i.e. that 
individual 
observations are not correlated. In contrast, true observations may contain 
a degree of correlation and this effect can interfere with periodogram 
statistics. To assure that the conclusions of this article are not affected
by the limitations of randomisation techniques, we compare our final results 
with semi-analytical estimates following \citet[][]{pressetal92-1} 
and \citet[][his Section 3.2]{paltani04-1}.

\section{The superhump signal}

To analyse the observations, we first subtract the linear trend and 
the offset from the visual magnitudes to get a distribution of 160 
values of $\Delta\,V$ with 
the mean value at zero as \citet{smak+waagen04-1} did.
One can then analyse the observed light curve using the AoV method and
the described randomisation techniques to determine the significance of the 
claimed superhump signal. However, simply analysing periodograms with 
constant trial periods is not sufficient in the context of superhumps as their
periods are usually not constant. Another important boundary condition 
for the numerical method is given by the considered range of periods. 
Using additional information e.g. derived from earlier observations of
superhumps may significantly affect the results. 
For example, if one can restrict the range of trial periods, 
a signal which has not been significant may become significant under the
condition given by the additional information.  
To analyse the statistical significance of the alleged superhump signal, 
we discuss pure periodogram analysis, time dependence, and the restriction to
a small range of trial periods in the following three Sections. 

\subsection{Pure periodogram analysis}

Using a rather broad range of constant trial periods corresponding to  
$7$\,c/d$\leq\,f\leq\,3$\,c/d, $N_b=5$, $N_c=2$, and 
$2\times10^4$ randomised datasets,
we find that the alleged superhump 
signal is not statistically significant, i.e. $p=0.275\pm0.003$.
This means that one finds a signal of the same significance in more than one 
forth of all 
light curves.  
This result reflects exactly the impression mentioned by 
\citet{pattersonetal05-1} that the evidence for the superhump signal
does not seem strong. 

%Our result of $p=0.275\pm0.003$ is in perfect agreement 
%with the estimates given in Sect.\,2 which confirms that our Monte-Carlo 
%approach correctly determines the statistical significance of periodicities 
%in uneven data sets. The distribution of $max(\Theta(P))$ for $N_b=5$ and 
%$N_c=2$ together with the obtained superhump signal is shown in the 
%top panel of Fig.\,\ref{fig_histb}.

\subsection{Time dependence}

In the previous Section we have shown that the AAVSO data  
does not contain a significant {\em{constant}} period. This does, 
however, {\em{not}} answer the question whether the superhump detection 
by \citet{smak+waagen04-1} is real or not because we used only constant 
trial periods, i.e. we ignored that 
superhump periods in general and the one claimed for U\,Gem in particular  
are time dependent. 
\citet{smak+waagen04-1} find that period changes
of the alleged superhump period are coherent and give a final fit for the
times of maxima (their Eq.\,(3)).
We calculate the value of the AoV statistic for this time-dependent
period and obtain $\Theta(P(t))=7.51$. Comparing this with the previously
obtained constant value $\Theta(P)=5.00$ shows that indeed the time-dependence
of the alleged superhump period increases the value of the statistic.
However, one has to take into account that the probability distribution
changes if one takes into account the time-dependence of the period.  
To determine the significance we need to compare the signal
with the distribution of $max(\Theta(P(t)))$ where $P(t)$ represents
every type of time-dependence acceptable for superhumps. We have incorporated
this in our Monte-Carlo simulations by using additionally time dependent trial
periods restricting ourself to coherent period changes. 
We used 
$P(t)=P_0+a\,\dot{P}$ where $1/3$\,d$\geq\,P_0\,\geq1/7$\,d and
$-0.0001\leq\,a\leq\,0.0001$. 
This is certainly reasonable as most superhump period derivatives and 
especially the one measured for U\,Gem are constant.
The effect of taking time dependence into account is illustrated 
in Fig.\,\ref{fig_fake}.
On the left hand side we used a broad range of trial periods and the alleged 
superhump signal is hardly distinguishable from noise. 
In other words, allowing for $dP/dt\neq\,0$ does not
significantly increase the statistical significance.
The bottom panel of Fig.\,\ref{fig_histb} shows the corresponding 
distribution of $max(\Theta(P(t)))$ derived from our high resolution 
Monte-Carlo simulations. 
The vertical lines indicate the $\Theta$-statistic 
required for 
$2\sigma$ and $3\sigma$ significance. Obviously, even the time-dependent
signal does not reach these values.   

\subsection{The range of periods}

\begin{figure}
\includegraphics[width=6.45cm, angle=270]{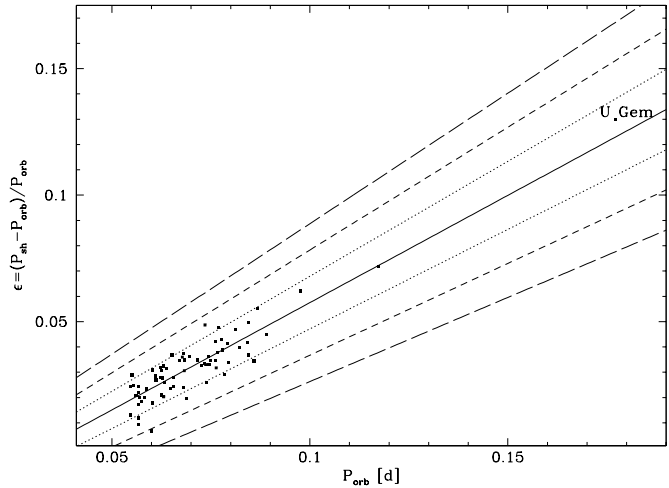}
\caption{\label{fig_eps} 
The $\epsilon-\Porb$ relation for SU\,UMa systems. The three dashed lines
represent the $1, 2, 3 \sigma$ regions around the linear fit.  
For the orbital period of U\,Gem the $3\sigma$ error of the 
linear fit corresponds to $4.83$\,c/d$\leq\,f\leq\,5.24$\,c/d.
}
\end{figure}

It has been realised by \citet{smak+waagen04-1} that the periodogram 
alone can not confirm the appearance of superhumps in U\,Gem. Therefore they
argue that the superhump period and the corresponding value of 
$\epsilon\equiv\,(P-\Porb)/\Porb=0.13$ are consistent with the mass ratio 
of U Gem and its long orbital period.
This led the authors to the reasonable interpretation that the ``observed'' 
signal represents a real superhump. 
Here, we test this argument using the numerical methods 
described above. 

\begin{figure}
\includegraphics[width=8.5cm, angle=0]{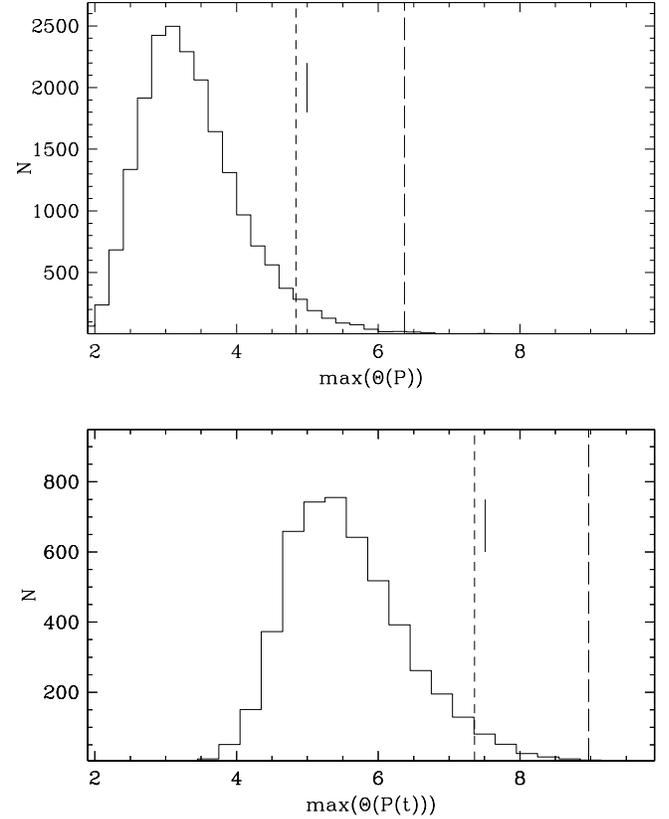}
\caption{\label{fig_hist3s} 
Distribution of the maximum of the statistic $\Theta(P)$ for 5 bins and
assuming that only periods in the $3\sigma$ interval of the $\epsilon-\Porb$
relation can be interpreted as superhumps. In the top panel we used only
constant periods while we took into account coherent time evolution for the
superhump periods in the bottom panel. 
The short solid vertical lines indicate the positions of the
claimed superhump signal. The dashed lines represent the $95\%$ and $99.7\%$ 
($3\,\sigma$) significance levels.  
In both cases the statistical significance of the superhump signal is
reaching $2\sigma$, i.e $p=0.0381\pm0.001$ (top panel) and $p=0.0384\pm0.003$ 
(bottom panel).  
}
\end{figure}

Fig.\,\ref{fig_eps} shows the $\epsilon-\Porb$ relation for SU\,UMa stars 
and the position of U\,Gem. Indeed, the claimed superhump signal lies 
directly on the extension of the linear relation for SU\,UMa systems. 
We approximated the $\epsilon-\Porb$ relation using linear regression 
(solid line in Fig.\,\ref{fig_eps}). The regions defined by the $1, 2, 3\,\sigma$ errors of the linear fit are also shown (dashed lines). 
Assuming that we know with $100\%$ confidence that 
the $\epsilon-\Porb$ relation has a linear extension up to long orbital 
period dwarf novae like U\,Gem, a new superhump signal should lie 
in the $3-\sigma$ region around the linear fit with a probability of 
$0.997$. Using this information means to analyse only those periods 
that are in agreement with the extension of the $\epsilon-\Porb$ relation,
i.e. $4.83$\,c/d$\leq\,f\leq\,5.24$\,c/d. 
In the case of time dependent periods, the effect is illustrated in 
Fig.\,\ref{fig_fake}. While the superhump signal does not produce a
significant peak in the $f$-$dP/dt$ plane if a broad range of trial periods
is used (left), it is clearly the highest peak in case the range of trialñ
pèperiods is restricted to periods in agreement with the $\epsilon-\Porb$
relation (right). 
This restriction indeed dramatically increases the significance 
of the ``observed'' signal as the distributions of $max(\Theta(P))$ 
and $max(\Theta(P(t)))$ are
shifted towards smaller values. 
The results of our detailed Monte-Carlo simulations are shown in 
Fig.\,\ref{fig_hist3s}. 
The $p$-value is below the $2\,\sigma$ significance level
in both cases, i.e $p=0.0381\pm0.001$ in the case of constant trial periods 
(top panel) and $p=0.0384\pm0.003$ when taking into account coherent period 
changes (bottom panel).

\subsection{Semi-analytical methods}

As mentioned in Section\,2, Monte-Carlo methods may fail if the analysed 
observations contain correlated data. 
To make sure that our results are not affected by this effect, we use 
semi-analytical methods to estimate the number of independent frequencies 
$M$. The power of the signal in the Lomb-Scargle periodogram 
(Fig.\,\ref{fig_ftaovpd}) and Eq.\,(1) then allow to derive a
statistical significance which can be compared with the results obtained by
our extended numerical simulations. 

Following \citet{pressetal92-1}, $M$ should not be very different from the
number of data points, i.e. $M\sim\,N_0=160$. 
This estimate for $M$ is, however, valid for the frequency range 
$0\leq\,f\leq\,f_N$ with $f_N$ being the Nyquist frequency. 
Assuming that $M$ decreases linearly with the 
frequency bandwidth \citep[see][for a discussion]{pressetal92-1}, 
the restriction to the small range of frequencies, 
i.e. $4.83$\,c/d$\leq\,f\leq\,5.24$\,c/d, leads to $M\sim30$. 
Using Eq.\,(1) and $z=6.74$ we derive
\begin{equation}
M\simeq\,30\rightarrow\,p\simeq\,0.035.
\end{equation}
This is obviously in perfect agreement with our
detailed Monte-Carlo simulations.

As an additional final (and less rough) test, 
the \citet{nemec+nemec85-1} method and the
\citet{pressetal92-1} estimate can 
be compared with the formalism recently proposed by 
\citet[][his Section 3.2]{paltani04-1}. 
For $n$ frequencies and an arbitrary threshold $\Theta^*$, 
this method requires $n\rm{Prob}(\Theta\geq\Theta^*)\ll\,1$. 
We follow Paltani in using $n(\rm{Prob}(\Theta\geq\Theta^*))=0.1$. 
With $n=1000$ frequencies in
the range of $4.83$\,c/d$\leq\,f\leq\,5.24$\,c/d, 
this requires $\rm{Prob}(\Theta\geq\Theta^*)=0.0001$ which (according to the
Fisher-Snedecor distribution for AoV statistics) results in $\Theta^*=6.3$.   
For $m=10.000$ randomised data sets we then determine the maximum value of 
$\Theta(f_{j=1,...,n})$ and the number of data sets with $max(\Theta(f_j))>\Theta^*$
gives an estimate for the number of independent frequencies $M$, 
i.e.   
\begin{equation}
M={\rm{\#}}({\rm{max}}(_{j=1,...,n}\Theta(f_j))>\Theta^*).
\end{equation}
Approximating
\#$({\rm{max}}(_{j=1,...,n}\Theta(f_j))>\Theta^*)$ with a Poisson distribution 
the error of $M$ is simply
$\Delta\,M=\sqrt{M}$ and we finally obtain
\begin{equation}
M=28\pm5\rightarrow\,p=\,0.033\pm\,0.006.
\end{equation}
This is again in very good agreement with our previous results. 

\section{The additional superhump}

\begin{figure}
\includegraphics[width=8.5cm, angle=0]{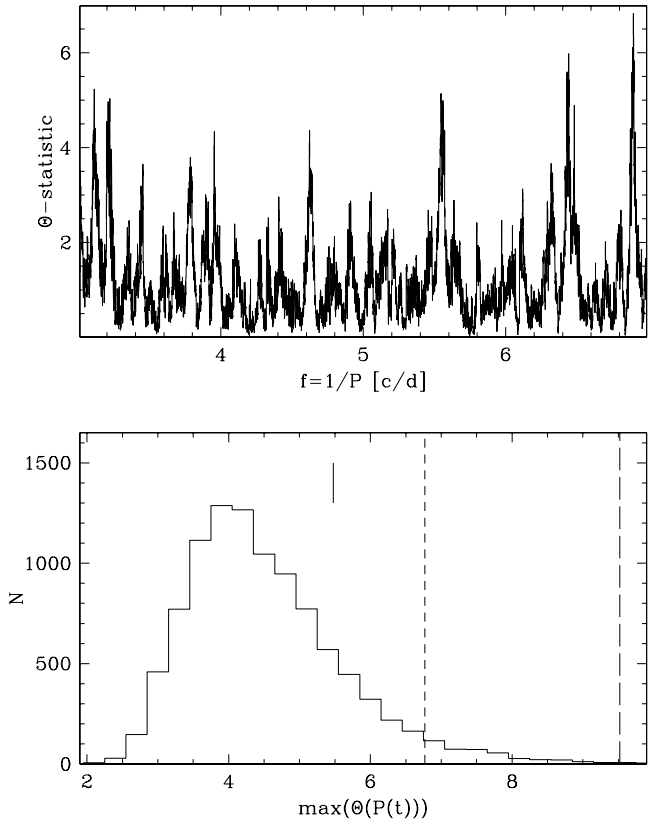}
\caption{\label{fig_add} 
AoV periodogram for the last 40 measured magnitudes covering
$JD=2446366-22122446382$ (top panel). The peak at $f\sim5.5$\,c/d has been 
interpreted as an ``additional superhump'' by \citet{smak+waagen04-1}. 
However, Analysing just the last 40 data points using our Monte-Carlo code
shows that this signal is not statistically significant. 
In the distribution of $max(\Theta(P))$ (bottom panel) the signal is far from 
reaching $2\sigma$ instead $p=0.20\pm0.02$.}
\end{figure}

In addition to the superhump signal at $f\sim5$\,c/d, 
\citet{smak+waagen04-1} 
claimed the existence of an additional superhump
around $f\sim5.55$ appearing during the final stages
of the outburst. To determine the significance of this signal
we analysed separately the last 40 data-points 
(covering $JD=2446366-22122446382$). 
Indeed, the strongest signal is no longer
the original superhump but the alleged additional superhump at
$f\sim\,5.55$. Using a broad range of periods
($3$\,c/d$\leq\,f\leq\,7$\,c/d) we obtain that this signal is even less 
significant than the alleged normal superhump, i.e. $p=0.38\pm0.02$. 
As this weak signal is the first additional superhump ever mentioned, and
hence no additional information like an established
relation between superhump excess and orbital period exists, 
the only restriction to the range of trial periods we can 
apply is to use only those periods that in
principal could be interpreted as superhumps, i.e.
$-0.076\leq\epsilon\leq\,0.241$. This corresponds
to $4.53$\,c/d$\leq\,f\,\leq\,6.11$\,c/d and we obtain a reduced p-value 
of $p=0.20\pm0.02$ which, however, is still far from representing 
a statistically significant value. The corresponding distribution 
of $max(\Theta(P))$ is shown in Fig.\,\ref{fig_add}.  

\section{Summary and conclusion}

We analysed the significance of the alleged superhump signal in the 
famous long 1985 outburst of U\,Gem using analysis of variance (AoV) and 
Monte-Carlo techniques. As randomisation methods may fail if the
observations are correlated, we derived in parallel semi-analytical 
estimates following \citet{pressetal92-1} and \citet{paltani04-1} which
perfectly confirm our Monte-Carlo results. The results of our analysis 
are:
\begin{itemize}

\item Using only the information provided by the 160 AAVSO data points of the 
1985 outburst of U\,Gem we do not find a statically significant periodicity. 
The probability to obtain a signal similar to the one detected by  
\citet{smak+waagen04-1} by chance is $p\gappr0.2$. 

\item If we restrict our numerical analysis to trial periods consistent with
the observed $\epsilon-\Porb$ relation for SU\,UMa systems we find that the
alleged superhump signal is statistically significant.    
The p-value decreases to $p\lappr0.04$ and the significance is above
$2\sigma$. In other words, the probability to 
detect by chance a periodic signal as strong as the observed one 
which is also consistent with the $\epsilon-\Porb$ is less than $0.04$.   

\item The alleged additional superhump \citep{smak+waagen04-1} is
statistically not significant, i.e. $p=0.20\pm0.02$. 

\end{itemize}

In addition to our detailed statistical analysis one should keep in mind that
the superhump signal is a typical one not only because of the agreement of the
$\epsilon-\Porb$ relation but also because of its (constant) amplitude, 
appearance 2-3 days after maximum, and its disappearance slightly before the
end of the outburst. In this sense, the determined statistical 
significance of $2\sigma$ can be considered as a lower limit. 
On the other hand, one should also be aware of the fact that extending
the observed linear correlation between superhump excess and orbital period 
towards longer orbital periods represents an {\em{assumption}} which is not
necessarily true.
However, balancing the pros and cons, we recommend to assume as the new 
working hypothesis that the mechanisms causing superhumps and 
superoutbursts in SU\,UMa systems probably also triggered the 1985 
superoutburst and -- keeping in mind the weak statistical significance -- 
superhumps in U\,Gem.
Concerning the triggering of superoutbursts the enhanced mass transfer
model is a very promising alternative to the TTI and 
its predictions are in agreement with the observations of U\,Gem 
\citep{lasota01-1,smak05-1}. In this scenario, the remaining big problem 
is a missing explanation for superhumps.

\begin{acknowledgements}
I thank the referee, Dr. Schwarzenberg-Czerny, for helpful comments and 
Dr. Tom Marsh for a brief but useful discussion during the European 
White Dwarf Workshop in Leicester. This work made use of data provided by the 
AAVSO. I acknowledge support by the Deutsches Zentrum f\"ur Luft-
und Raumfahrt (DLR) GmbH under contract No. FKZ 50 OR 0404 and 
FONDECYT (grant 1061199).
\end{acknowledgements}

%\bibliography{../../aamnem99,../../aabib}
%\bibliographystyle{../../apj}

%\begin{thebibliography}{30}
%\expandafter\ifx\csname natexlab\endcsname\relax\def\natexlab#1{#1}\fi

%\bibliography{../aamnem99,../aabib}
%\bibliographystyle{../apj}

\end{document}